# Generic Properties of Random Gene Regulatory Networks

(Running title: Small motifs determine the dynamics of large networks)

Zhiyuan Li[1,2], Simone Bianco[2], Zhaoyang Zhang[3], and Chao Tang[1,*]

[1] Center for Quantitative Biology, School of Physics and Peking-Tsinghua Center for Life Sciences, Peking University, Beijing 100871, China

[2] Department of Bioengineering and Therapeutic Sciences, University of California, San Francisco, California 94158, USA

[3] Department of Physics, Beijing Normal University, Beijing, 100871, China

[*] Correspondence: tangc@pku.edu.cn.

Keywords: Genetic network, dynamic attractors, complex systems

## Abstract

Modeling gene regulatory networks (GRNs) is an important topic in systems biology. Although there has been much work focusing on various specific systems, the generic behavior of GRNs with continuous variables is still elusive. In particular, it is not clear typically how attractors partition among the three types of orbits: steady state, periodic and chaotic, and how the dynamical properties change with network's topological characteristics. In this work, we first investigated these questions in random GRNs with different network sizes, connectivity, fraction of inhibitory links and transcription regulation rules. Then we searched for the core motifs that govern the dynamic behavior of large GRNs. We show that the stability of a random GRN is typically governed by a few embedding motifs of small sizes, and therefore can in general be understood in the context of these short motifs. Our results provide insights for the study and design of genetic networks.

## INTRODUCTION

Gene regulatory network as a dynamical system has long attracted much attention [1-3]. As a product of evolution, a biological network possesses certain "evolved" properties and features that are specific to the function it is performing [4-6]. Nonetheless, it would be revealing to study and understand the "generic" or typical properties of the networks of the kind, to serve as a "null" model that can be compared and contrasted with the various specific natural networks. Kauffman and others have studied random Boolean networks and a key finding is that these networks exhibit an order to chaos transition as the network connectivity is increased [7, 8]. Thus, for a typical random Boolean network of connectivity larger than two, the typical behavior of the network dynamics is chaotic. This result seems at odds with the observation that most, if not all, natural gene regulatory networks have highly ordered dynamics [9]. Less attention has been paid to generic properties of GRNs modeled by continuous variables, which represent an important class of modeling. Glass and colleagues studied a piecewise linear GRN model with a hybrid rule: continuous time but Boolean-like regulations [10]. They found that the probability of observing different types of attractors changes with network size and connectivity: the probability of reaching periodic or chaotic attractors first increases with network size and then decrease. In this work, we model GRNs by ordinary differential equations (ODEs) and with biological-like regulation rules. We first systematically explore their dynamical properties with different network sizes, connectivity, regulation rules and fraction of inhibitory links. We further investigate the general mechanism that determines network stability. We found that generally the dynamic behavior of a GRN is governed by short motifs, so that the probability for a GRN to have steady state or non-stationary behavior can be predicted by analyzing these short motifs, even for large GRNs.

**RESULTS**

**The model**

In characterizing the network topology, $N$ stands for the number of nodes (genes), $K$ stands for connectivity, and $\varepsilon$ is the fraction of inhibitory regulations. For three-node networks ($N$=3), we enumerated all possible topologies. For larger networks, regulations between pairs of genes were randomly assigned to present with probability $\frac{K}{N}$, and to be either inhibition or activation with probability $\varepsilon$ and 1-$\varepsilon$, respectively. Once assigned, the regulation links are fixed. Note that a gene can simultaneously activate some of its targets and inhibit the others, and that a gene can be regulated by multiple genes. The time derivative of the $k$-th gene product is generally written as:

$$\frac{dg_k}{dt} = \frac{1}{\tau_k}(G_k(g_{a1}, g_{a2}, ..., g_{i1}, g_{i2}, ...) - g_k). \quad (1)$$

The generation rate of the gene product, $G_k$, is a function of all the activators $(g_{a1}, g_{a2},...)$ and the inhibitors $(g_{i1}, g_{i2},...)$ that regulate the *k*-th gene. We model gene regulations with Hill functions. Each activating regulation is modeled as $\frac{g_a^n}{g_a^n + K_a^n}$, and inhibitory regulation as $\frac{K_i^n}{g_i^n + K_i^n}$. When a gene is regulated by two or more genes, it is necessary to specify the regulation "logic", i.e. the effect of the multiple regulations [11, 12] (Fig.S1). The regulation logic depends on the promoter structure and the interactions between and among the transcription factors (TFs), the promoter and the RNA polymerase. Three rules commonly used in modeling are presented here in the main text: the "AND" rule (multiplying all the activating and the inhibiting Hill functions), the "Additive" rule (summing up these Hill functions) and the "Strong inhibition" rule (first summing up the activating Hill functions and then multiply the sum with all inhibiting Hill functions) [13, 14]. Results on other rules are shown in Figure S3.

**Small networks**

In exploring the attractor properties of random GRNs with different topological characteristics, we first studied small networks – all network topologies made of three nodes [15, 16]. Here, different transcription regulatory rules yield similar results: mono-stable and bi-stable asymptotic states are the most probable outcomes, appearing with probabilities around 70% and 25%, respectively (Supplementary Table S1-3 and Fig.S2 ). Periodic limit-cycles and chaos are infrequent (less than 1%), which is consistent with a recent work by Zhang et al. [17]. Multi-stability is uncommon. The highest number of stable states is $2^3=8$ and the only core motif to have 8 fixed points is the topology with self-activation on all nodes but no inter-gene regulations (Fig. S2).

**Large networks**

Next, we extended our study to larger networks with equal fraction of activation and inhibition. Previous studies have shown that Boolean networks exhibit an order to chaos transition when connectivity exceeds a threshold [8, 18]. Using ODEs, we investigated the dynamic properties of networks with 100 nodes and increasing connectivity. As summarized in Figure 1, the results depend largely on the transcription regulatory rules. When the gene regulation logic is "AND" or "Additive", a slightly higher occurrence of limit-cycles is observed for K smaller than 5 (Fig. 1(a) and 1(b)). Further increase of connectivity results in most trajectories being at steady state. This situation is comparable to the piecewise linear model, where the basin size of the asymptotic periodic and chaotic states increase and then decreases with connectivity [10]. However, if the rule is chosen to be "Strong inhibition", the percentage of periodic or chaotic trajectories increases drastically with network connectivity (Fig. 1(c)). When the network is very densely connected ($K$=20), more than 30% of the trajectories are periodic and around 40% are chaotic. This increased instability is comparable to the Random Boolean Networks. Multi-stability is suggested to be essential for cell fate decisions during development and differentiation [19]. Interestingly, under "Strong Inhibition" rule, though the total number of all types of attractors increases with network size, the number of steady states decreases (Fig. 1(d)), implying either biological systems with multiple cell fates are carefully rewired to avoid random behaviors, or posses multi-oscillator rather than multi-steady as the fate attractor. These results suggest that the stability of large GRNs depends on the rule of transcription regulation, and may become more chaotic in certain cases.

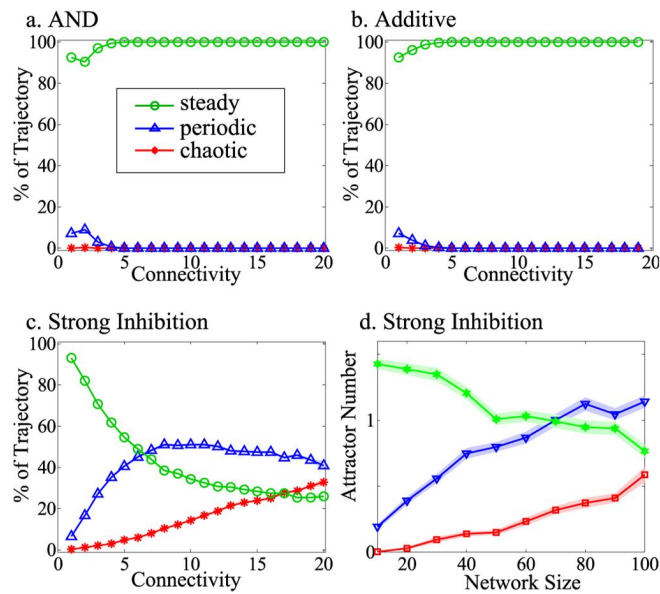

Figure1. (A)–(C) The percentages of asymptotically steady state (green circle), periodic (blue triangle), and chaotic (red star) trajectories with increasing network connectivity ($K$=1–20), in 100-node GRNs. The regulation rule is shown above the panel. (D) The number of three types of attractors (green star : steady state; blue triangle: limit cycle; red square: chaotic attractor) with increasing network size under Strong Inhibition rule ($K$= 10). The results of each panel are averages from $10^3$ random GRNs with equal fractions of activating and inhibiting links. Each set of ODEs (corresponding to one network topology) is simulated with 100 random initial conditions.

**The role of inhibitory links in network stability**

We now proceed to study the influence of inhibitory links on network dynamics. We varied the fraction of inhibition links in randomly generated GRNs from 0 to 1. In large networks ($N$=100, $K$=5), the percentages of trajectories entering periodic (Fig. 2(a)) and chaotic (Fig. 2(b)) attractors increase under the "AND" rule and the "Strong inhibition" rule. In more densely connected networks with $N$=100 and $K$=10, when all the interactions are inhibitory, over 50% of the trajectories show asymptotically chaotic behavior (Fig. S6). This phenomenon that inhibition destabilizes network is mainly contributed by the increased number of the all-inhibition three-node and five-node negative feedback loops (Fig. 2(c-d)), as will be discussed in detail later.

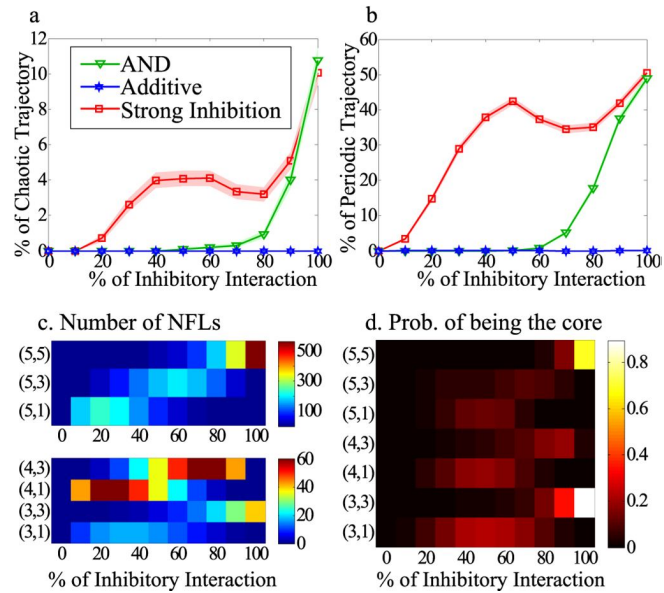

Figure2. The percentage of chaotic (A) and periodic (B) trajectories with increasing fraction f of inhibitory interactions under different regulation rules (green: AND; blue: Additive; red: Strong Inhibition). (C)The total number of negative feedback loops and (D) the average occurrence of the core motifs, with increasing fraction of the inhibitory interaction. Numbers in brackets specify the length of, and the number of inhibitions in the loop, respectively. (N= 100, K=5, Strong Inhibition rule).

**Core motifs**

Next, in order to understand the mechanism that determines network stability, we identified the "sources" of unstable behaviors in GRNs. For a network entering periodic or chaotic attractors, we "freeze" links one by one by assigning the regulating node its average value. If this action turned the attractor into a stable one or significantly perturbed the shape of the attractor (defined by 10% change in period for limit-cycle or 5% change in amplitude for chaos), the link is marked as "essential". We then identified the core motifs constituted by these essential links (see Fig.S7 for an example). For a large network ($N$=100), the number of the core motifs is small and their lengths are short; the majority of them are negative feedback loops. We denote negative feedback loops of lengths between 3 to

5 as "short motifs" (See Fig.S8 for topologies of all 7 short motifs). For "AND" and "Additive" rules with $K=2$, there is on average 1 feedback loop that is the core of the dynamics, comparing to more than $10^8$ loops in total (Fig. 3(a)). Over 80% of the core motifs are short motifs. For "Strong Inhibition" rule with $K=5$, there are on average 2 feedback loops that are essential for the network's dynamical behavior, comparing to more than $10^{33}$ loops in total (Fig. 3(b)). 60% of the core motifs are short motifs, and 81% of all networks have at least one such short core motif. In order to test if the presence of these motifs is sufficient to drive instability in the whole network, we freeze all links in the network but those that are part of the essential loops. In 85% of all cases the attractors remain to be periodic or chaotic, suggesting these motifs are the most likely sources of instability.

We argue that the reason for short motifs being the main source of limit-cycle is that faster oscillators override the slower ones. In all cases where a negative feedback loop is being identified as the core motif for the limit-cycle, there is a positive correlation between the period of the limit-cycle and the length of the loop (Fig. 3(c)). Shorter motifs in general have faster frequency. In freezing links of a network entering a limit cycle, with 0.7% chance the network enters another limit cycle with period changed more than 10%. More than 70 % of these cases have their period increased, implying that a longer negative feedback loop is unlikely to drive the limit-cycle unless the shorter core motif is disrupted.

Negative feedback loops of lengths three to five constitute the majority of core motifs (Figs. 3(a-b)). Given that all the links outside the core circuits can be frozen without affecting much of the dynamics, we treat all outside regulations as a constant "environment" that exerts a constant regulation on each node within the core motif. This environment can be decomposed into two parts: a multiplicator $m$, and a summand $s$. For example, if gene A is negatively regulated by gene B in the core motif, then

$$\frac{d[A]}{dt} = \frac{1}{\tau_A}(s + m \cdot \frac{K_A^n}{K_A^n + [B]^n} - [A]), \qquad (2)$$

where $s$ and $m$ are constants, the values of which are determined by other frozen non-core regulations (if exist) to A.

As shown in Fig. 3(d), higher m and lower s lead to higher probability of limit-cycle for short motifs (measured by the fraction of randomly sampled parameter sets entering limit cycle attractors). In other words, the less influence from the environment, the higher the probability of limit-cycle.

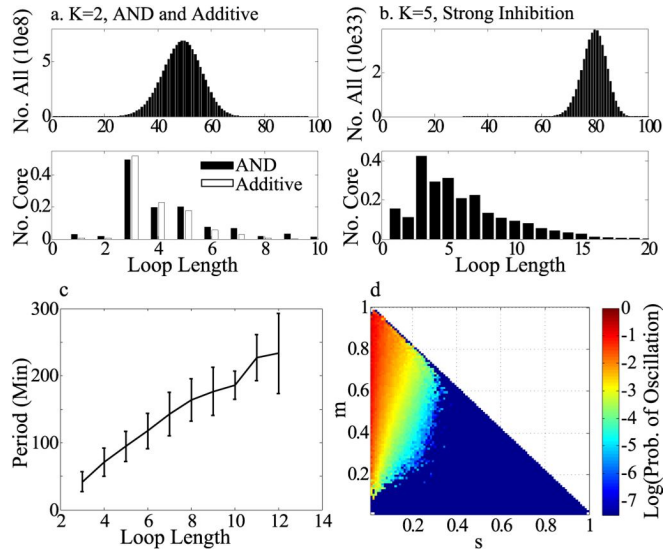

Figure3. (A)–(B) The length distribution of all loops (upper panels) in a network compared to that of the core motifs (lower panels) for AND and Additive rules (A) and Strong Inhibition rule (B). (C) The relation between the limit cycle period and the core loop length. (D)Relation between the environmental parameters $m$ and $s$,and the probability of limit cycle for short negative feedback loops (coded by color). Due to normalization, $m+s < 1$.

**Source of instability**

The connection between network stability and topological characteristics can be understood through the total occurrence of short negative feedback motifs in the network, and their probability to oscillate due to the environmental parameters $m$ and $s$. For example, the fact that inhibition destabilizes network (Figs. 2(a-b)) is mainly contributed by the increased number of the all-inhibition three-node and five-node negative feedback loops as the fraction of inhibitory links increases (Figs. 2(c-d)). Another example is the different stability of networks with the same $N$ and $K$ but different transcription regulatory rules. Their total numbers of various motifs are identical, but the environmental parameters for the short motifs distribute in distinct region of the $m$ - $s$ plane. In the case of "AND" rule, $s$ is zero but $m$ is in general very small (Fig.4 (a)); for the "Additive rule", $m$ and $s$ tend to distribute in the lower right region, making these motifs less likely to become the source of limit-cycle (Fig. 4(b)). With "Strong inhibition" rule, $m$ and $s$ can reach the left corner of the $m$ - $s$ plane where the probability of limit-cycle is the highest (Fig. 4(c)). Finally, the relation between network connectivity and stability can be qualitatively predicted by the trade-off between the number of short motifs and their ability to oscillate. When the connectivity increases, the total number of short motifs increases (Fig. 4(d)). However, the probability for each loop to oscillate drops due to increased interference from the environment (Fig. 4(e)). In the case of "Strong Inhibition" rule, the increase in number for short motifs transcends the decrease in their limit-cycle probability, resulting in a rising trend of limit-cycle and chaos for the whole network (Fig. 4(f)). In the case of "AND" or "Additive" rule, the probability of limit-cycle drops too fast to be compensated by the increase of motif number, resulting in networks in high connectivity dominated by steady state behavior (Fig. 4(f)). The percentage of non-stationary trajectories (limit-cycle and chaos) predicted by short motifs agrees well with the simulation results (Fig. 1 (a-c) and Fig. 4(f)).

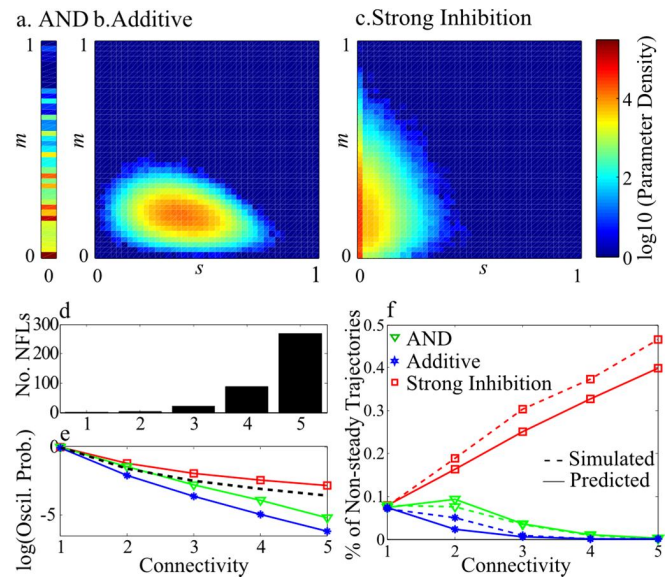

Figure4. The distribution of m and s for all short motifs under different rules (A) AND; (B) Additive; (C) Strong Inhibition; (D) The number of short motifs with increasing connectivity. (E)The probability of limit cycle for short motifs with increasing connectivity obtained from the map of Figure3D. Dash line specifies1/Number of short motifs. (F)The percentage of non-steady state trajectories with increasing connectivity predicted by multiplying results in (D) and (E), compared to that obtained by simulation. (N=100)

**DISCUSSION**

In summary, we investigated the properties of attractor landscape for random gene regulatory networks. Studies of generic network behaviors may provide insights on the selection forces acting on real biological networks. Two counteracting forces shape the biological network as we see it: evolutionary pressure selects for specific topologies that optimize the desired biological function, while random drift pushes the network towards a more non-organized structure. Our results on large GRNs suggest that gene networks are typically stable under several transcription regulatory rules and inhibitor fractions. Thus, in the evolution of gene networks and during the execution of the various network functions, nature does not have to pay much attention to keep the network dynamics well behaved. On the other hand, chaos and limit-cycles do occur and their occurrence increases with the fraction of inhibitory regulations. In the *E. coli* transcriptional network, there are about twice as many activators as inhibitors [20]. The reason might be related to the overall stability of the network. Moreover, we have shown that in large networks the increase of connectivity may or may not lead to instability, depending on the regulation logic. This would suggest that biological networks may adjust the regulation logic to achieve desirable dynamic properties.

Another interesting result of ours is that network dynamics is typically governed by small core motifs. It is known that in biological oscillators, such as the cell cycle [21] and the circadian clock [22, 23], although the limit-cycle can involve a large number of genes and proteins, the drivers of the limit-cycle are usually consisted of a small number of genes and proteins [24, 25]. We confirmed that our results also hold for the piecewise linear models [26, 27] (Fig.S10-12). It would be interesting to further investigate the role of small core motifs in other types of networks [25, 28].

SUPPLEMENTARY MATERIALS

The supplementary materials can be found online with this article at DOI10.1007/s40484-014-0026-6.

ACKNOWLEDGEMENTS

This work was supported in part by NSF (DMR-0804183; CMMI-0941355), NIH (R01 GM097115; P50 GM081879), MOST (2009CB918500) and National Natural Science Foundation of China. We thank Leon Glass and Gang Hu for helpful discussions.